\documentclass[prb,floats,aps,superscriptaddress,showpacs]{revtex4}
\usepackage{amsmath}
\usepackage{graphicx} 
\usepackage{epsfig}

\newcommand{\elle}{{\it l}}
\begin{document}
\date{\today}

\author{Crist\'obal L\'opez}
\affiliation{
Instituto
Mediterr\'aneo de Estudios Avanzados IMEDEA (CSIC-UIB),
     Campus de la Universidad de las Islas  Baleares,
E-07122 Palma de Mallorca, Spain.
}
\author{Umberto Marini Bettolo Marconi} 
\affiliation{
Dipartimento di Fisica, Via Madonna delle Carceri, 68032 Camerino (MC), Italy
}

\title{Multiple time-scale approach for a system of
Brownian particles in a non-uniform temperature field.}
\begin{abstract}

The Smoluchowsky equation for a system of interacting
Brownian particles in a temperature gradient is derived from
the Kramers equation by means of a multiple time-scale 
method. The interparticle interactions are assumed to be represented
by a mean-field description.
We present numerical results that compare well with
the theoretical prediction together with an extensive discussion on the 
prescription of the Langevin equation in  overdamped systems.
\end{abstract}

\maketitle

\section{Introduction}

The erratic motion of tiny particles suspended in a fluid of lighter 
particles is a known phenomenon, going under the name of Brownian motion,
 of fundamental interest  in Statistical 
Mechanics~\cite{Gardiner,Risken,Vankampen}.
A classical method to study this physical process
is the Langevin equation, where the
influence of the solvent particles on the heavier particles is modeled 
by two force terms. One term represents a viscous force,
with friction coefficient $\gamma$, the other a stochastic force, whose 
intensity depends on the temperature of the solvent.
Alternatively, the description of the dynamics of a system of Brownian
particles in an external field
may rest upon the Fokker-Planck  equation (FPE) for
the probability density of the particles. 
The two levels of description, namely, 
a) the {\it Klein-Kramers} equation~\cite{Kramers}, which
considers the evolution of the probability density of position
and velocity, and b) the {\it Smoluchowski}
equation~\cite{Smoluchowski}, for the
probability distribution of position only, have
deserved a great deal of attention.
The passage from the first description to the second one is well understood
for the case of ideal non-interacting particles immersed in a 
heat bath at constant temperature.
In fact, as shown by Wilemski  and Titulaer~\cite{Titulaer}, the 
derivation of the Smoluchowski equation
from the Kramers-Klein equation,
can be performed
by means of a perturbation expansion in terms of 
the inverse of the friction coefficient, $\gamma$.
On the other side, one can derive the Smoluchowski equation more directly
after taking the overdamped
limit $\gamma \to \infty$ in the Langevin equation
and neglecting the inertial acceleration term. 
Physically speaking, this procedure corresponds to the fact that
the velocities of the particles thermalize rapidly, so that
only the probability distribution of positions of the particles matters.

Thermal gradients, besides causing
a heat flow, can induce a mass flow in systems such as colloids.
The phenomenon is termed thermodiffusion, thermophoresis or Ludwig-Soret 
effect~\cite{Ludwig,Soret,Landau,Degroot,Dhont,Parola,Bringuier} 
and is relevant
because it represents a tool to manipulate and concentrate molecules in
solution. Thermophoresis, in fact, typically enhances the concentration
in colder regions. Recently, it has been employed to form two dimensional
colloidal crystals using $2 \mu m$ polystyrene beads. Such a
method leads to applications in microfiltration, particle accumulation
and molecular detection on surfaces~\cite{Duhr}.
Nevertheless, in spite of its technological
interest, not many studies provide a  derivation of the governing
equation for the concentration field in the presence of a temperature
gradient externally imposed.
A problem arises when one considers the $\gamma \to \infty$ limit:
the resulting stochastic equation contains a multiplicative
noise term (the temperature is space-dependent).
This in turn means that the associated
FPE depends on the particular prescription used to derive it,
i.e. one encounters the so called Ito-Stratonovich dilemma.
To solve it, Matsuo and Sasa~\cite{Sasa} recently obtained
the Smoluchowski
equation from a perturbation expansion in powers of the inverse
of $\gamma$. Two important messages arise from their work: first,
the underdamped dynamics is free from the Ito-Stratonovich dilemma,
and, second,  their perturbation scheme is unbounded in time
and one needs a renormalization treatment  to make it convergent.

The main objective of this work is to present
a systematic and convergent
derivation of the Smoluchowski equation for 
a system of interacting Brownian particles in a temperature gradient. 
The interaction is of mean-field type, so that our starting point,
the Kramers equation, can be written down in close form for the
one-particle distribution function.
We employ a multiple time-scale method
which has been designed to deal 
with non-uniformities in systems with more than a time scale.
The approach is  quite appropriate 
for our objectives, 
since in the  large friction limit there is a very
clear time-scale separation in the system:  in a very short period
the velocities of the particles relax to 
their thermal equilibrium value, whereas on a
much longer time scale also the positions of the particles
assume configurations compatible with their steady state distribution.
In this regard, the multiple time-scale method has been used
to perform the passage from the Kramers to the Smoluchowski
equation for an ideal gas at constant temperature 
as an expansion in powers of the small parameter $\gamma^{-1}$,
providing a uniformly valid result~\cite{Bocquet1,Bocquet2}. 
More recently, the case of 
interacting hard-core particles in isothermal systems  has also been studied
within this framework~\cite{Umberto}.

Another important part of this work is the discussion
about the  Ito versus Stratonovich description of the system.
This is based on numerical results of the 
Langevin dynamics.

The plan of this paper is the following:  in section
\ref{section:FP} we present the full Langevin equation,
its overdamped limit, and the corresponding
Kramers and Smoluchowski 
descriptions of the system.  
In section \ref{section:MS}   the derivation of the Smoluchowski 
equation
with the multiple time-scale method is presented. In 
section \ref{section:numerics} we present the numerical results, 
with some discussion on the different Ito-Stratonovich prescriptions,
and in section \ref{section:conclusions} we give
our conclusions.

\section{Microscopic model. Smoluchowski  and Kramers equations.}
\label{section:FP}

Let us consider a system of $N$ interacting Brownian particles 
of mass $m$ in an external force field. For the sake of simplicity, 
let us assume  only one spatial 
dimension. The system is  also
under the influence of an external non-uniform temperature
field so that the dynamics of the particles is given by:

\begin{eqnarray}
&\frac{d x_i }{dt}  &= v_i, \nonumber \\
& m \frac{d v_i}{dt}& = -m \gamma v_i +f_{ext}(x_i)
+\sum_{j\neq i}f_{int}(x_i-x_j)+\sqrt{T(x_i)}\xi_i(t),
\label{langevin}
\end{eqnarray}
for $i=1,...,N$, and
where $\gamma$ is the friction coefficient,
$T$ is a non-uniform {\it heat-bath temperature},
$f_{ext}$ is an external force, $f_{int}$ is the
interaction force between pair of particles, 
 and
 $\xi_i(t)$
is a Gaussian white noise with zero average and 
correlation 
\begin{equation}
\langle \xi_i(t)\xi_j(s) \rangle  = 2 \gamma m k_B  \delta_{ij} \delta(t-s).
\end{equation} 
 $\langle \cdot \rangle$ indicates 
the average over a statistical ensemble of realizations,
and $k_B$ is the Boltzmann constant.
To study the dynamics of the system we consider the single particle 
probability distribution function, $P(x_i,v_i,t)$, of the position and velocity
variables. Its evolution is given by 
the Fokker-Planck equation~\cite{Gardiner} 
also known as Kramers equation:

\begin{equation}
\frac{\partial}{\partial t}P(x,v,t)+
\Bigr [v \partial_x + \frac{f_{ext}(x)+f_m(x)}{m}\partial_v \Bigr] P(x,v,t) =
\gamma \partial_v\Bigr[ v P(x,v,t)  
+\frac{k_B T(x)}{m} \partial_v P(x,v,t)\Bigr].
\label{kramers}
\end{equation}

Here $f_m(x)$ is the molecular field which takes the form
\begin{equation}
f_m(x)=\int dx' f_{int}(|x-x'|) \rho(x')=
-\int dx' \frac{d U(|x-x'|)}{dx'} \rho(x'),
\label{fm}
\end{equation}
being $U(|x-x'|)$ the pair potential energy (i.e. $f_{int} (x_i -x_j)=
\frac{\partial U (x_i- x_j)}{\partial x_i}$), 
and $\rho(x)=\int dv P(x,v,t)$. It is very important to realize under what
conditions on the interparticle interactions one can 
write down 
eq.~(\ref{kramers}) (see~\cite{Umberto}). 
Note that  with no approximations the 
Kramers-FPE  for interacting particles for the one-particle distribution
function, $P(x,v,t)$, depends on the two-particle distribution
$P_2(x,v,x',v',t)$. 
The main hypothesis done in Eq.~(\ref{kramers})
 is that whenever $U(x-x')$ is a smooth function
of particle separation
one may assume a mean-field approximation  for the two-particle
distribution function $P_2 (x,v,x',v',t)\approx P (x,v,t) P(x',v',t)$. 
In the following we shall define
$f=f_{ext}+f_m$.
 
Since the temperature $T(x)$ in
Eq.~(\ref{kramers}) multiplies  a derivative with respect to $v$,
one realizes that the Kramers equation is free from the so called
Ito-Stratonovich dilemma. On the contrary, when one considers 
the overdamped limit, $\gamma \to \infty$, in
Eq.~(\ref{langevin}) one obtains the following Langevin equation:
\begin{equation}
\frac{d x_i }{dt}  = \frac{f(x_i)}{m \gamma} + 
\frac{\sqrt{T (x_i)}}{m \gamma}\xi_i(t),
\label{langevinoverdamped}
\end{equation}
which leads to different equations for the 
distribution of the spatial position  
(the so-called Smoluchowski equation)
depending on the prescription adopted to 
integrate the stochastic evolution.
In fact, the corresponding FPE for Ito
and Stratonovich conventions, the two rules usually applied,  
are respectively:
\begin{eqnarray}
&\frac{\partial}{\partial t}P(x,t)
=-\frac{1}{m \gamma}\partial_x (f P)+ \frac{k_B}{\sqrt{m}
\gamma}\partial_x^2 (T P),&
\label{itosmo} \\
&\frac{\partial}{\partial t}P(x,t)
=-\frac{1}{m \gamma}\partial_x (f P)+ \frac{k_B}{\sqrt{m}
\gamma}\partial_x (T^{1/2}
\partial_x(T^{1/2} P)).&
\label{stratosmo}
\end{eqnarray}

The aim of this work is to derive 
 a Smoluchowski equation for 
the system using the multiple time-scale method,
via an expansion in the inverse of the friction coefficient.  
Our starting point is the Kramers equation, Eq.~(\ref{kramers}).
At this stage it is convenient to introduce 
the following dimensionless variables: 
\begin{equation}
\tau\equiv t\frac{v_{T}}{\elle}, \qquad V\equiv\frac{v}{v_{T}}, 
\qquad X\equiv\frac{x}{\elle},\qquad \Gamma=\gamma\frac{\elle}{v_{T}},
\label{adim1}
\end{equation}

\begin{equation}
F(X)\equiv\frac{\elle f(x)}{m v_T^2},\qquad 
\tilde P(X,V,\tau)\equiv\elle v_T P(x,v,t),
\label{adim2}
\end{equation}
being $v_{T_0}=\sqrt{k_B T_0/m}$ the thermal velocity
at the reference temperature  $T_0$ (which can be taken
equal to $1$), 
and $\elle$ is a typical length scale of the system, for example
the effective radius of the particles. The non-dimensional
Kramers equation for $\tilde P(X,V,\tau)$ 
denoting, $\tilde T (X) = T (x)$,
is:
\begin{equation}
\frac{\partial  \tilde P(X,V,\tau)}{\partial \tau}
 =\Gamma \partial_V (V \tilde P)
+ \Gamma \tilde T (X) \partial_V^2 \tilde P - V \partial_X \tilde P
-F \partial_V \tilde P.
\label{nondimkramers}
\end{equation}

In the following we shall skip 
the tildes in our notation.
The (non-dimensional) Kramers equation, Eq.~(\ref{nondimkramers}),
is our definitive starting point and in the next section
we proceed to present the  set of eigenfunctions
that will be used in the subsequent section when 
we make  use of the multiple time-scale  approach.

\section{Method of solution}
\label{section:MS}

Standard solutions of Eq.~(\ref{nondimkramers}) go through an expansion
in a basis set of eigenfunctions. 
First,
it is convenient to separate the spatial dependence from the 
velocity dependence  of the probability distribution.
We rewrite Eq.~(\ref{nondimkramers}) as:  
\begin{equation}
\Gamma L_{FP} P(X,V,\tau)=
\Bigr [\frac{\partial  }{\partial \tau}+
V \frac{\partial }{\partial X}+
F\frac{\partial }{\partial V}\Bigr ]P(X,V,\tau),
\label{evolv}
\end{equation}
where we introduced the {\it Fokker-Planck} operator
\begin{equation}
L_{FP} P(X,V,\tau) =\frac{\partial}{\partial V}\Bigl[
T(X)\frac{\partial }{\partial V }+V\Bigl]   P(X,V,\tau).
\label{fokkerp}
\end{equation}

We will expand $P$ in terms of the eigenfunctions $H_{\nu}(X,V)$ of
$L_{FP}$~\cite{Sasa} which are related to the Hermite polynomials:
\begin{equation}
L_{FP}H_{\nu}(X,V)=-\nu H_{\nu}(X,V)
\end{equation}
with
\begin{equation}
H_{\nu}(X,V)\equiv \frac{T(X)^{\nu/2} }{\sqrt{2\pi T(X)}}
(-1)^{\nu} \frac{\partial^{\nu}}{\partial V^{\nu}} 
\exp(-\frac{V^2}{2 T(X)}).
\label{basis}
\end{equation}

We now express the solutions to Eq.~(\ref{evolv}) as
a linear combination of products 
of the type $\phi_{\nu}(X,\tau) H_{\nu}(X,V)$ and
after some algebra obtain:
\begin{eqnarray}
\sum_{\nu=0}^\infty\Bigl\{
\Bigl[&&\frac{\partial \phi_{\nu}(X,\tau)}{\partial \tau}
+\Gamma \nu\phi_{\nu}(X,\tau)\Big] H_\nu(X,V)+
\\\nonumber
&&\sqrt{T(X)}\Bigl[\frac{\partial \phi_\nu(X,\tau)}{\partial X}
-\frac{F(X)}{T(X)}\phi_\nu(X,\tau) \Bigl]H_{\nu+1}(X,V)+\sqrt{T(X)}
\nu \frac{\partial \phi_\nu(X,\tau)}
{\partial X}(\delta_{\nu,0}-1)H_{\nu-1}(X,V) \\\nonumber
&&+\frac{T'(X)}{2\sqrt{T(X)}}\phi_\nu(X,\tau)[H_{\nu+3}(X,V)
+2(\nu+1)H_{\nu+1}(X,V)+\nu H_{\nu-1}(X,V)]\Bigl\}=0
\label{brinkmanhiera}
\end{eqnarray}
where the prime denotes derivation with respect to the argument.
Equating the coefficients of the same basis functions, $H_{\nu}$,
we obtain an infinite  hierarchy of equations relating the various moments
$\phi_{\nu}(X,\tau)$. Such a hierarchy can be truncated by
the multiple scale method, as shown hereafter.


\subsection{Multiple time-scale analysis}

In the multiple time-scale analysis one determines the temporal evolution
of the distribution function $P(X,V,\tau)$ in the regime
$\Gamma^{-1}<<1$ by means of a perturbative method. 

In order to construct the solution one replaces the single
physical time scale, $\tau$, by a series of auxiliary time scales
($\tau_0,\tau_1,..,\tau_n$) which are related to the original variable
by the relations $\tau_n=\Gamma^{-n}\tau$. Also the original
time-dependent function, $ P(X,V,\tau)$,  
is replaced by an auxiliary function, $ P_a(X,V,\tau_0,\tau_1,..)$,  
depending on the $\tau_n$, which are treated as independent variables.
Once the equations corresponding to the various orders have been 
determined, one returns to the original time variable and to the
original distribution.

We begin by replacing the time derivative
with respect to $\tau$ by a sum of partial derivatives:
\begin{equation}
\frac{\partial}{\partial \tau}=\frac{\partial}{\partial \tau_0}
+\frac{1}{\Gamma} \frac{\partial}{\partial \tau_1}
+\frac{1}{\Gamma^2} \frac{\partial}{\partial \tau_2}+...
\label{mult}
\end{equation}
The auxiliary function, $ P_a(X,V,\tau_0,\tau_1,...)$,
is expanded as a series of powers of $\Gamma^{-1}$
\begin{equation} 
P_a(X,V,\tau_0,\tau_1,\tau_2,...)=
\sum_{s=0}^{\infty} \frac{1}{\Gamma^s} 
P_a^{(s)}(X,V,\tau_0,\tau_1,\tau_2,...).
\label{pn}
\end{equation}

Each term $P_a^{(s)}$ is then projected over the eigenfunctions $H_{\nu}$,
\begin{equation}
P^{(s)}_a(X,V,\tau_0,\tau_1,...)=\sum_{\nu=0}^{\infty}  
\psi^{(s)}_{\nu}(X,\tau_0,\tau_1,\tau_2,...)H_{\nu}(X,V).
\label{phi}
\end{equation}

At this stage one substitutes the time derivative~(\ref{mult})
and expressions~(\ref{pn})-(\ref{phi}) into Eq. 
(\ref{nondimkramers}), and identifying terms 
of the same order in $\Gamma^{-1}$ in the equations
one obtains a hierarchy of relations
between the amplitudes $\psi^{(s)}_{\nu}$. The advantage
of the method over the naive perturbation theory is that 
secular divergences can be eliminated at each order of perturbation
theory and thus uniform convergence is achieved.

The most important function in all this expansion is $\psi^{(0)}_{0}$.
Its physical meaning becomes clear if one integrates over $V$ 
the auxiliary density probability function 
$P_a(X,V,\tau_0,\tau_1,\tau_2,...)$. Assuming, as in the standard expansion,
that $\psi^{(\nu)}_{0}=0$ for $\nu \ge 1$ 
one sees that $\psi^{(0)}_{0}$ is the spatial density probability
function. 
We shall consider the partial derivative,
$\partial_{\tau_i} \psi^{(0)}_{0}$, and obtain the 
evolution equation for
$\psi^{(0)}_{0}$ to order $\Gamma^{-i}$.
For the sake of completeness 
we briefly show how the method works.  
To order $\Gamma^0$ one finds:
\begin{equation}
L_{FP} \Bigr[\sum_\nu \psi^{(0)}_{\nu} H_\nu\Bigr]=0,
\label{g0}
\end{equation} 
and concludes that only the amplitude $\psi^{(0)}_{0}$ is non-zero. 


We consider, next, terms of order  $\Gamma^{-1}$ and write 
(note that on the l.h.s we always have the application of the 
Fokker-Planck operator to the auxiliary function to the order
of interest, in this case $s=1$) :
\begin{eqnarray}
&
-(\psi^{(1)}_{1} H_1 
+2 \psi^{(1)}_{2}  H_2 
+3 \psi^{(1)}_{3}  H_3 +\sum_{\nu \ge 3} \nu \psi^{(1)}_{\nu} H_{\nu} ) = 
H_0 
\partial_{\tau_0} \psi^{(0)}_{0}
 & \\ \nonumber
&+\sqrt{T(x)} H_1 [\partial_X \psi^{(0)}_{0} 
- \frac{F}{ T} \psi^{(0)}_{0}
+\frac{ T' }{T}
\psi^{(0)}_{0} ]
+\frac{T'}{2\sqrt{T}} H_3  \psi^{(0)}_{0}.
&
\label{g1}
\end{eqnarray}

By equating the coefficients multiplying  $H_{0}$ 
we find the relation:
\begin{equation}
\frac{\partial \psi^{(0)}_{0} }{\partial \tau_0}=0,
\label{psi0ta}
\end{equation} 
i.e. the amplitude $\psi^{(0)}_{0}$ is not a function of  $\tau_0$.
We also obtain for the amplitude $\psi^{(1)}_{1}$ the equations 
\begin{equation}
\psi^{(1)}_{1}=\Bigl[
-\sqrt{T(X)} \partial_X \psi^{(0)}_{0}
+ \frac{F}{\sqrt{T(X)}} \psi^{(0)}_{0}
-\frac{T'(X)}{\sqrt{T(X)}\psi^{(0)}_{0}}
\Bigl].
\label{psi0t}
\end{equation} 

The remaining  amplitudes are
\begin{equation}
\psi^{(1)}_{2}=0,
\qquad \psi^{(1)}_{3}=-\frac{1}{6}\frac{T'(X)}{\sqrt{T(X)}}\psi^{(0)}_{0},
\qquad \psi^{(1)}_{\nu}=0 \ \  if  \ \ \ \nu \ge 3.
\label{psi0txx}
\end{equation}
By iterating the method we obtain the correction
of order $\Gamma^{-1}$ to the evolution equation of $\psi^{(0)}_{0}$,
which reads

\begin{equation}
\partial_{\tau_1} \psi^{(0)}_{0}
=-\partial_X (\sqrt{T}\psi^{(1)}_{1}),
\label{g3}
\end{equation}
and using Eq.~(\ref{psi0t}) we can rewrite it as:
\begin{equation}
\partial_{\tau_1} \psi^{(0)}_{0}=
\partial_X \Bigl[\partial_X (T \psi^{(0)}_{0})
- F \psi^{(0)}_{0}\Bigl]
\label{psifinal}
\end{equation}

Finally, we recover the original time variable, $\tau$,
and obtain the most important result of this work, that is,
the evolution equation for the spatial probability
density:
\begin{equation}
\partial_{\tau} \psi^{(0)}_{0}= \frac{1}{\Gamma}
\partial_X \Bigl[\partial_X (T \psi^{(0)}_{0})
- F \psi^{(0)}_{0}\Bigl] + O(1/\Gamma^2).
\label{final}
\end{equation}
With $O(1/\Gamma^2)$ we mean terms of 
or $1/\Gamma^2$ and superior.
The first comment that arises from Eq.~(\ref{final}) 
is that it is of the Ito type if one uses this prescription 
for the overdamped Langevin equation. However, here we did not
use any prescription for the Langevin equation, and 
Eq.~(\ref{final}) was derived from an expansion of the 
Kramers equation. Moreover, the generalization of 
Eq.~(\ref{final}) to arbitrary spatial
dimensions is straightforward.

 Using a more usual notation for the spatial density probability function,
and following the line of reasoning of Marconi and Tarazona~\cite{Tarazona},
one could write the following governing equation
\begin{equation}
\frac{\partial \rho(x,t)}{\partial t}=
\nabla\Bigr \{\nabla D(x)\rho(x,t)+\frac{1}{m\gamma}\rho(x,t) \nabla\bigr[ 
\frac{\delta {\cal F}_{ni}[\rho]}{\delta \rho(x,t)} + 
V_{ext}(x)\bigr]\Bigr \},
\label{fm2}
\end{equation} 
where $D(x)=k_B T(x)/m\gamma$ and ${\cal F}_{ni}$ is the 
excess over the ideal gas contribution to the free energy~\cite{Archer}.
The second term renormalizes the effective diffusion constant of the
particles and may become important at higher concentrations.

In the next section we proceed to check numerically our results
and to enter in a possible discussion about the Ito-Stratonovich
dilemma.

\section{Numerical results}
\label{section:numerics}

This section is devoted to numerically check our analytical results.
The way to proceed is to compare the results of  numerical simulations
of  a system of $N$ particles following the dynamics given by
Eqs.~(\ref{langevin}) or rather Eq.~(\ref{langevinoverdamped}).
We shall compare
these with the analytical solution, if available, of the corresponding
density equation: Eq.~(\ref{final}) for the first case,
or the Smoluchowski in the Ito and Stratonovich prescriptions,
Eqs.~(\ref{itosmo}-\ref{stratosmo}),
for the second case.

In order to obtain a simple
analytical solution of the density equation 
we assume a vanishing external force and molecular field,
$F=0$, particles
have  mass one, the system is one-dimensional and particles 
positions restrict to the interval $[0,1]$. We also assume that 
the temperature field has a linear spatial dependence
$T(X)= T_0 + T_1 X $, with $T_0$ and $T_1$  positive constants. 
Finally, we impose  zero flux  boundary conditions of the system 
 at the extremes of the spatial 
interval, i.e., there are infinite walls at these extremes where
the particles rebound and cannot scape from the system.

The equation for the probability density that 
we have derived from the Kramers equation with the
multiscale approach in an expansion in $\Gamma$, Eq.~(\ref{final}),  is now
(for aesthetic reasons we denote in the following 
$\psi^{(0)}_{0} (X,\tau)=
 \rho (X,\tau)$)
\begin{equation}
\partial_{\tau}  \rho (X,\tau)= \frac{1}{\Gamma}
\partial^2_X (T(X)  \rho)=
-\partial_X J(X),
\label{densnumerica}
\end{equation}
where $J$ is the flux. The boundary conditions read $J(0)=J(L)=0$.
The  stationary  solution is given by:
\begin{equation}
\rho^s(X) = \frac{N T_1}{(T_0 + T_1 X)\ln{[(T_0 + T_1 L)/T_0]}},
\label{estacionaria}
\end{equation}
where we have used the normalization condition
$\int_0^L dX \rho^s (X)=N$.

On the other side,
in the overdamped limit one has 
to distinguish between the two prescriptions. In this paper we discuss 
the two more standard: Ito and Stratonovich. The Ito-Smoluchowski equation
is also given by Eq.~(\ref{densnumerica}) and the stationary
solution is obviously given by Eq.~(\ref{estacionaria}). A very
important difference here is that it has not been derived
as an expansion of inverse powers of $\Gamma$, and thus it
is exact and no higher order corrections should appear.
In the Stratonovich 
prescription the density equation, which is also exact and not 
an approximation to order  $\Gamma^{-1}$, is given by:
\begin{equation}
\partial_{\tau}  \rho (X,\tau)= \frac{1}{\Gamma}
\partial_X (T^{1/2} \partial_X
 T^{1/2}  \rho).
\label{densnumericastrato}
\end{equation}
With the same boundary conditions the stationary Stratonovich solution is
given by:
\begin{equation}
\rho^s_{strat}=\frac{N T_1}{(T_0 + T_1 X)^{1/2}
[2 (\sqrt{T_0 + T_1 L}-\sqrt{T_0})]}
\label{estacionariastrato}
\end{equation}

\begin{figure}
\begin{center}
\epsfig{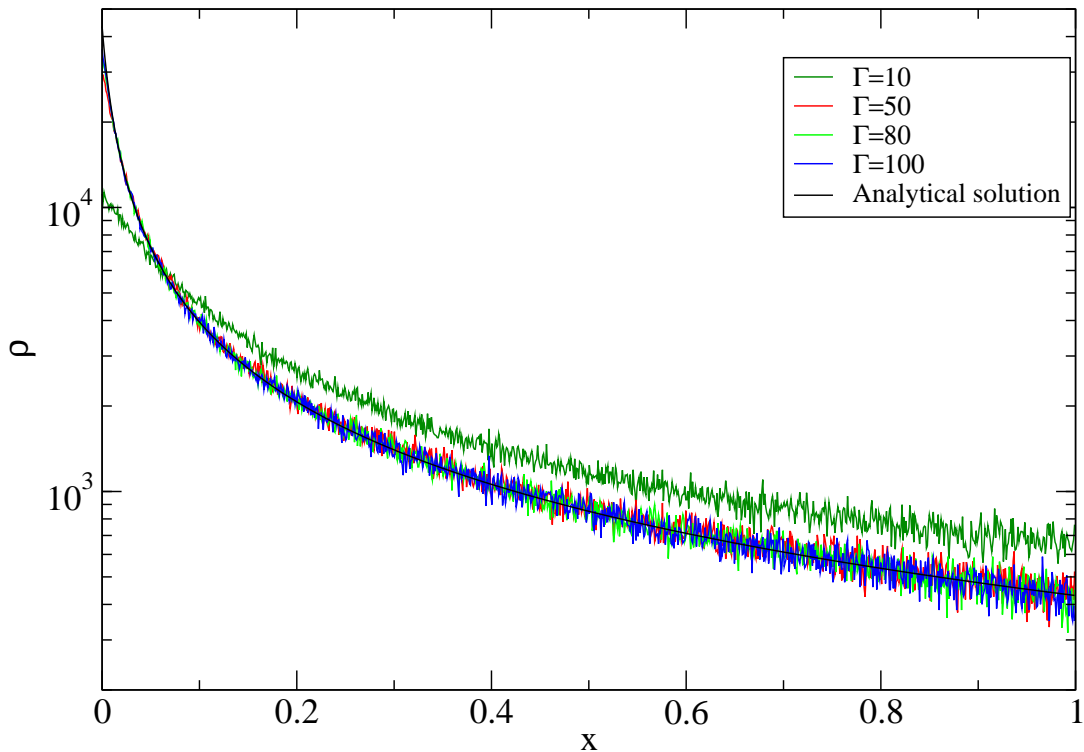}
\end{center}
\caption{(color online) Plot of the stationary density of particles for different
values of $\Gamma$. The particle dynamics correspond to the Langevin
equation Eq.~(\ref{langevin}), and the density of particles has been
computed by making a coarse-graining in the particle positions. With
the solid black line we plot the analytical solution Eq.~\ref{estacionaria},
that fits well for large $\Gamma$ values. The other parameters used
in the simulation are $dt=0.0001$ and the cell size for
the coarse-graining $dx=1/1000$.
}
\label{fig:densidad}
\end{figure}

Now we show the numerical results for the Langevin particle dynamics.
We begin with the underdamped dynamics,  and consider
$2000$ particles evolving according to Eq.~(\ref{langevin}). We take
$T_0=0.1$ and  $T_1=10$
and all our measurements are performed when the system is 
stationary. Average values are obtained by sampling  
$200$ different realizations.
 A stationary density of particles is defined
at every spatial point  by
counting the number of particles in
every cell of size $dx=1/1000$.   The time step used
is
$dt=0.0001$.  
For different values of $\Gamma$ the results are shown in
Fig.~\ref{fig:densidad}. One can see that for large values
of $\Gamma$ the comparison with the analytical result,
Eq.~(\ref{estacionaria}), is excellent. 
Obviously, the smaller the value of $\Gamma$ the larger
the corrections that we are neglecting and the
agreement worsens.

\begin{figure}
\begin{center}
\epsfig{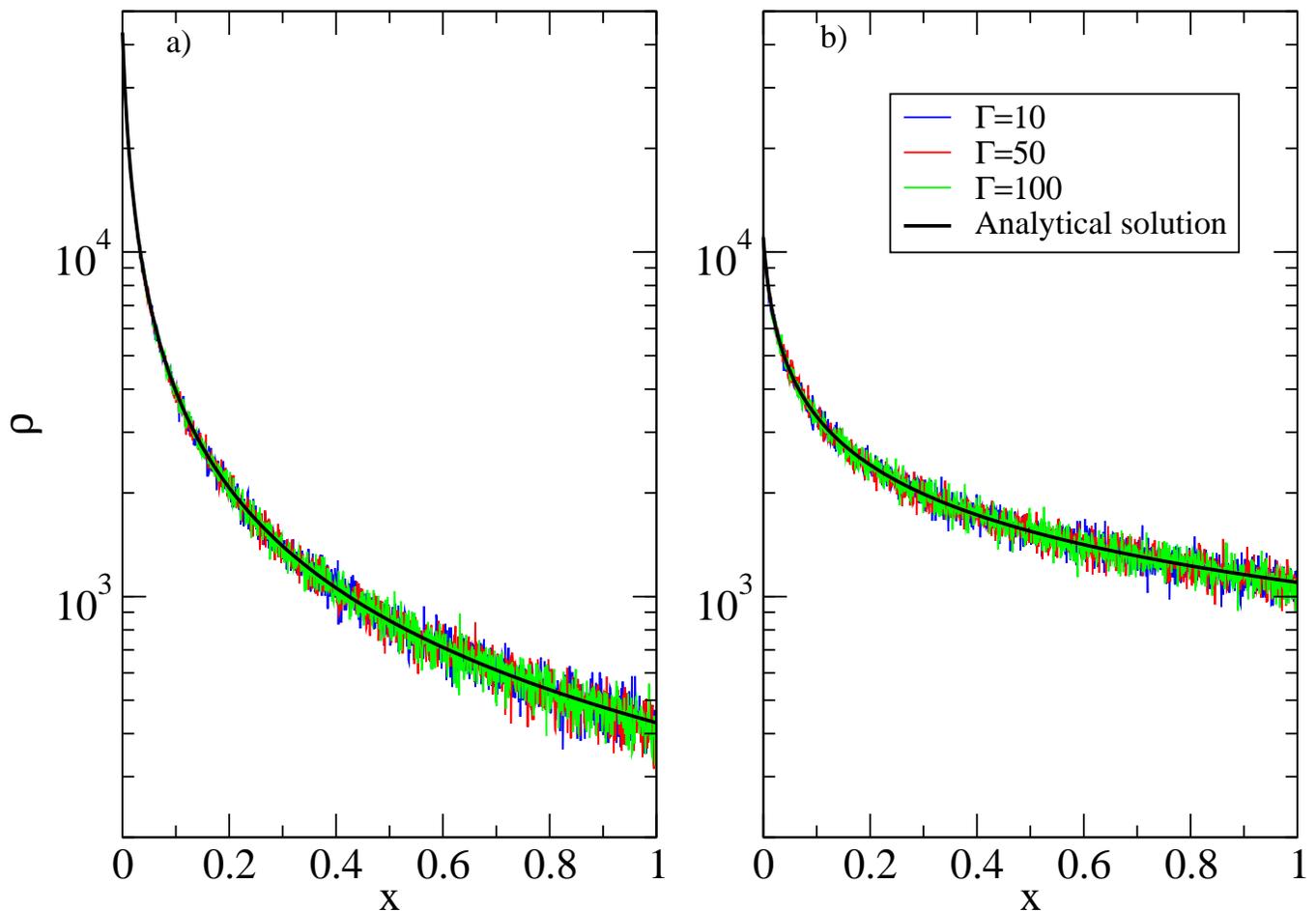}
\end{center}
\caption{(color online)
Coarse-grained density field obtained from the overdamped
Langevin dynamics Eq.~\ref{langevinoverdamped}. Left panel corresponds
to the Ito interpretation of this equation, and right to the
Stratonovich one. We take different values of the parameter $\Gamma$
as indicated in the legend box.
With the solid line we plot the stationary density solutions as 
given by Eqs.~\ref{estacionaria} and \ref{densnumericastrato},
for Ito and Stratonovich respectively.
The rest of the parameters used in the simulation are 
$dt=0.0001$ and the cell size to make the coarse-graining of the
position of the particles is $dx=1/1000$.
}
\label{fig:stratonovich}
\end{figure}

In  Fig.~\ref{fig:stratonovich} we show the results obtained 
from the simulations of the overdamped dynamics, 
Eq.~(\ref{langevinoverdamped}), where
we have also coarse-grained the distribution of particles to
compute a density of particles. Interpreting it like Ito the results
are in perfect agreement with the   Ito-Smoluchowski
stationary solution, Eq.(\ref{estacionaria}). This is shown in the
left panel of the figure for different values of $\Gamma$. 
 When it is understood in the Stratonovich
calculus, for which we have to add a new term 
$\frac{T'(X)}{2 m \gamma}$ in the r.h.s. 
of Eq.~(\ref{langevinoverdamped}) to make the numerical simulation,
 again the results coincide with
the stationary density solution in the Stratonovich
calculus, Eq.~(\ref{estacionariastrato}), as shown in the 
right panel of Fig.~\ref{fig:stratonovich}  for
various values  $\Gamma$.
A very important difference encountered in these results
with respect to those in Fig.~\ref{fig:densidad} is their
independence with respect to the value of $\Gamma$: for 
any of its values the agreement with the stationary density
is always excellent. In fact, one can realize that
once the limit has been performed, the value of $\Gamma$ only
renormalises the time scale of the problem.

\section{Conclusions}
\label{section:conclusions}
In the present paper, we have derived
the form of the governing equation for 
the probability density of interacting Brownian particles in the case of an
inhomogeneous medium, whose temperature  varies in space.
The interparticle interactions have been treated in a mean-field
framework.
 We have shown, 
using the Klein-Kramers equation associated to the fully inertial
stochastic dynamics, that,
to leading order, one obtains a Smoluchowski equation for
the particle distribution which has the same form as the
equation obtained by applying the Ito prescription to the
overdamped dynamics.

\section{Acknowledgments}

C.L. acknowledges financial support from MEC (Spain) and FEDER through project
CONOCE2 (FIS2004-00953)
and from the bilateral project Spain-Italy HI2004-0144.
 He also acknowledges a 
{\sl
Ram{\'o}n y Cajal} research fellow of the Spanish MEC.
U.M.B.M. acknowledges a grant COFIN-MIUR 2005, 2005027808.

\end{document}